\documentclass{article}

\usepackage{arxiv}

\usepackage[utf8]{inputenc} 
\usepackage[T1]{fontenc}    
\usepackage{hyperref}       
\usepackage{url}            
\usepackage{booktabs}       
\usepackage{amsfonts}       
\usepackage{nicefrac}       
\usepackage{microtype}      
\usepackage{lipsum}		
\usepackage{graphicx}
\usepackage{doi}
\usepackage{graphicx}

\title{32-bit RISC-V CPU Core on Logisim}

\author{ \href{https://orcid.org/0009-0001-4383-0579}{\includegraphics[scale=0.06]{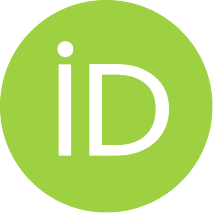}\hspace{1mm}Siddesh D. Patil} \\
	Department of Electrical Engineering\\
	Veermata Jijabai Technological Institute (VJTI)\\
	Mumbai, India\\
	\texttt{sdpatil\_b20@el.vjti.ac.in } \\
	\And
	\href{https://orcid.org/0000-0000-0000-0000}{\includegraphics[scale=0.06]{orcid.pdf}\hspace{1mm}Premraj V. Jadhav} \\
	Department of Electrical Engineering\\
	Veermata Jijabai Technological Institute (VJTI)\\
	Mumbai, India\\
	\texttt{pvjadhav\_b20@el.vjti.ac.in } \\
    \And
    \href{https://orcid.org/0000-0000-0000-0000}{\includegraphics[scale=0.06]{orcid.pdf}\hspace{1mm}Siddharth Sankhe} \\
	Department of Mechanical Engineering\\
	Veermata Jijabai Technological Institute (VJTI)\\
	Mumbai, India\\
	\texttt{sssankhe\_b20@me.vjti.ac.in} \\
}

\renewcommand{\headeright}{}
\renewcommand{\undertitle}{}

\hypersetup{
pdftitle={32-BIT RISC-V CPU CORE ON LOGISIM},
pdfsubject={q-bio.NC, q-bio.QM},
pdfauthor={Siddesh D. Patil, Premraj V. Jadhav},
pdfkeywords={RISC-V, cpu, logisim},
}

\begin{document}
\maketitle

\begin{abstract}
	This project focuses on making a RISC-V CPU Core using the Logisim software\cite{b1}\cite{b2}\cite{b3}. RISC-V is significant because it will allow smaller device manufacturers to build hardware
without paying royalties and allow developers and researchers to design and experiment with a
proven and freely available instruction set architecture. RISC-V is ideal for a variety of
applications from IOTs to Embedded systems such as disks, CPUs, Calculators, SOCs, etc. RISC-V(Reduced Instruction Set Architecture) is an open standard instruction set architecture (ISA) based on established reduced instruction set computer (RISC) principles. Unlike most other ISA designs, the RISC-V ISA is provided under open source licenses that do not require fees to use

\end{abstract}

\keywords{RISC-V \and CPU \and Logisim \and 32-bit}

\section{Introduction}
To build our CPU Core, we need to assemble 6 components mainly:
\begin{itemize}
    \item 1. PC logic
    \item 2. Register File
    \item 3. I-Mem
    \item 4. D-Mem
    \item 5. ALU
    \item 6. Decode Logic
\end{itemize}
    On integrating these components , we get a single pipelined CPU Core, our project heavily lies
on the CAO ( Computer Architecture and Organization) Domain. This project not only focuses on
making CPU Core using pre-existing components modules but also making each and every
small component from Flip Flops to Adders.
Originally developed by researchers at the University of California, Berkeley\cite{b4}, starting in 2010,
RISC-V represents one of many instruction set architectures (ISAs) that allow programmers and
the software they write to directly control computer hardware. The open-source flexibility of
RISC-V has made it an increasingly popular chip architecture for companies such as computing
storage giants Seagate and Western Digital Corp., China’s e-commerce giant Alibaba, along with
government initiatives backed by the U.S. military’s Defense Advanced Research Projects
Agency (DARPA). Our Cpu focuses on RV32I Instruction Set architecture, RV32I contains 40 unique instructions,
though a simple implementation might cover the ECALL/EBREAK instructions with a single
SYSTEM hardware instruction that always traps and might be able to implement the FENCE
instruction as a NOP, reducing base instruction count to 38 total.

\begin{figure}
    \centering
    \includegraphics[width=0.8\textwidth]{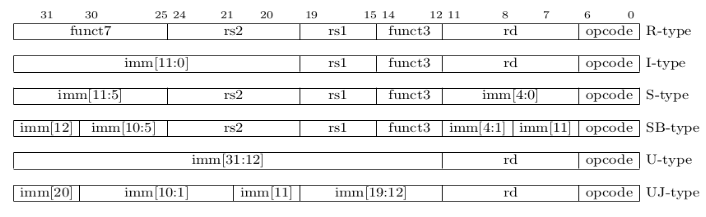}
    \caption{RISC-V BASE instruction formats showing immediate variants.}
    \label{fig:enter-label}
\end{figure}

\section{PROCEDURES AND RESULTS:}
\textbf{Logisim :}
Logisim is an educational tool for designing and simulating digital
logic circuits. With its simple toolbar interface and simulation of
circuits as you build them, it is simple enough to facilitate learning
the most basic concepts related to logic circuits. With the capacity
to build larger circuits from smaller subcircuits, and to draw
bundles of wires with a single mouse drag, Logisim can be used
(and is used) to design and simulate entire CPUs for educational
purposes.

\begin{figure}
    \centering
    \includegraphics[width=0.8\linewidth]{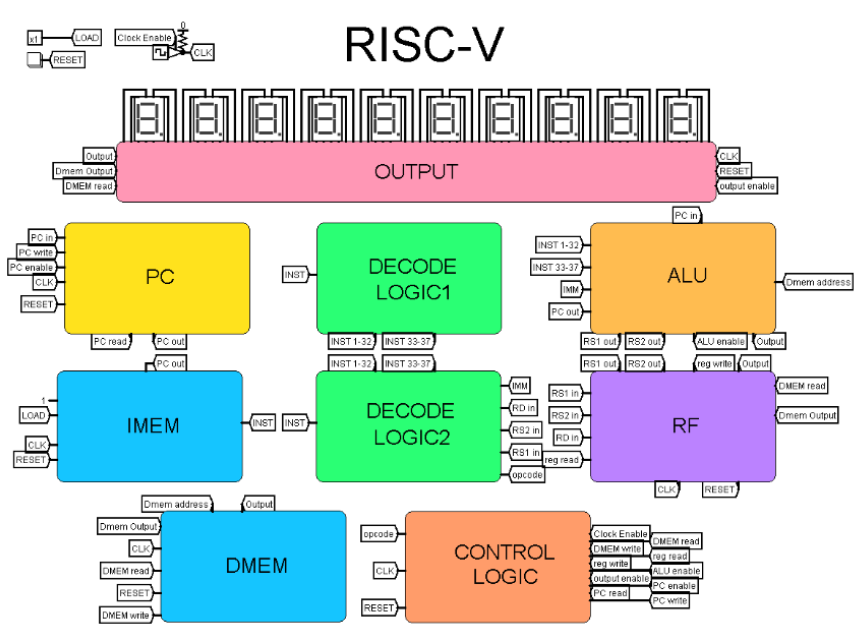}
    \caption{Logisim Circuit of the RISC-V CPU Core}
    \label{fig:enter-label}
\end{figure}

\begin{figure}
    \centering
    \includegraphics[width=0.8\linewidth]{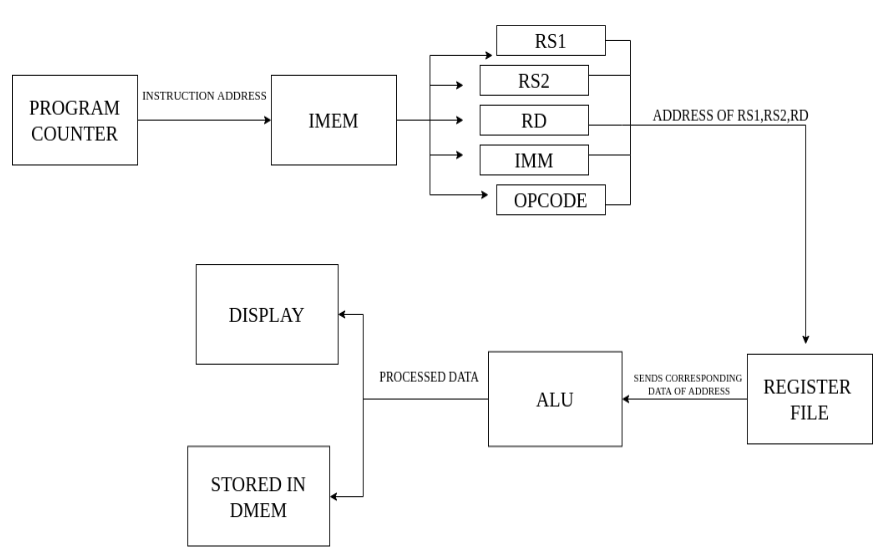}
    \caption{Block Diagram of RISC-V CPU Core}
    \label{fig:enter-label}
\end{figure}

\subsection{Fundamental Building Blocks}

\subsubsection{Clock}
The clock toggles its output value on a regular schedule as long as ticks are enabled. The clock's
cycle can be configured using its High Duration and Low Duration attributes. Clock is an
important component of our circuit as it can be used to induce delay and while making a
pipelined circuit, it helps us.
Note that Logisim's simulation of clocks is quite unrealistic: In real circuits, multiple clocks will
drift from one another and will never move in lockstep. But in Logisim, all clocks experience ticks
at the same rate.

Clocks are important components and are used to make Flip Flops and Registers. The delay
induced by the clocks helps to keep the signal from intermixing in Adders, that's why we prefer
Carry Lookahead adder over Ripple Carry Adder.

\subsubsection{Flip Flops}
Flip-flop or latch is a circuit that has two stable states and can be used to store state
information – a bistable multivibrator. The circuit can be made to change state by signals
applied to one or more control inputs and will have one or two outputs. It is the basic storage
element in sequential logic. Flip-flops and latches are fundamental building blocks of digital
electronics systems used in computers, communications, and many other types of systems. Flip
flops are used as storage devices and store 1 bit binary data ,one of its two states represents a
"one" and the other represents a "zero". Such data storage can be used for storage of state, and
such a circuit is described as sequential logic in electronics.
For this project, D flip flop (fig 4) was used for the Memory device.

\begin{itemize}
\item D flipflop is a modified version of SR flipflop in which compliment of set signal is provided as reset. It avoids same values of set and reset which can cause error. 
\end{itemize}

\begin{figure}
    \centering
    \includegraphics[width=0.5\linewidth]{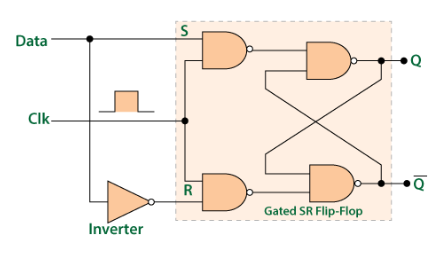}
    \caption{D-Flipflop}
    \label{fig:enter-label}
\end{figure}

\subsubsection{Adders}
An adder is a digital circuit that performs addition of numbers. In many computers and other
kinds of processors adders are used in the arithmetic logic units or ALU. They are also used in
other parts of the processor, where they are used to calculate addresses, table indices,
increment and decrement operators and similar operations. We have half adder and full adder
circuits.
\begin{itemize}
    \item Half Adder: 
A half adder can add two bits and give output as sum and carry.

Disadvantages: It doesn’t have any carry input pin so these adders cannot be extended to
perform addition of multiple bits.

    \item Full Adder: 
A full adder can add all three bits ( two operands and 1 carry bit). It outputs sum and carry based
on the inputs. Multiple full adders can be combined to add multiple bits.

\end{itemize}

\subsubsection{Multiplexers (Mux)}
A multiplexer is used to select between two or more inputs. The select lines identify the input to drive to the output. A mux can have $2^n$ input lines for n select lines. Mux always has only one
output line.A multiplexer makes it possible for several input signals to share one device or
resource, for example, one analog-to-digital converter or one communications transmission
medium, instead of having one device per input signal. Multiplexers can also be used to
implement Boolean functions of multiple variables.

\textit{\textbf{images}}

\subsection{Registers}
An electronic register(fig 5) is a form of memory that uses a series of flip-flops to store the individual
bits of a binary word, such as a byte (8 bits) of data. The length of the stored binary word
depends on the number of flip-flops that make up the register.

The binary word to be stored is applied to the four D inputs and is remembered by the flip-flops
at the rising edge of the next clock (CK) pulse. The stored data can then be read from the Q
outputs at any time, as long as power is maintained, or until a change of data on the D inputs is
stored by a further clock pulse, which overwrites the previous data.
Types of Register are:
\begin{itemize}
    \item Processor
    \item RAM
    \item Data register
    \item PC (Program counter)
    \item IMem
\end{itemize}

\begin{figure}
    \centering
    \includegraphics[width=0.5\linewidth]{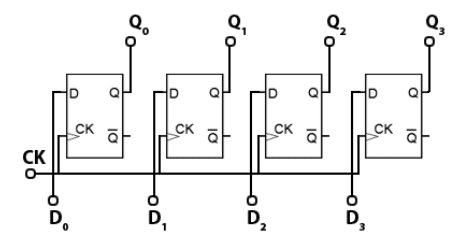}
    \caption{A simple 4 bit register}
    \label{fig:enter-label}
\end{figure}

\subsection{Processor Core}

\subsubsection{ALU \& Control Flow unit}

An arithmetic-logic unit (ALU) (fig 6,7 and 8) is the part of a computer processor (CPU) that carries out
arithmetic and logic operations on the operands in computer instruction words. In some
processors, the ALU is divided into two units, an arithmetic unit (AU) and a logic unit (LU).

\begin{figure}
  \centering
  \begin{minipage}{0.6\textwidth}
    \includegraphics[width=\linewidth]{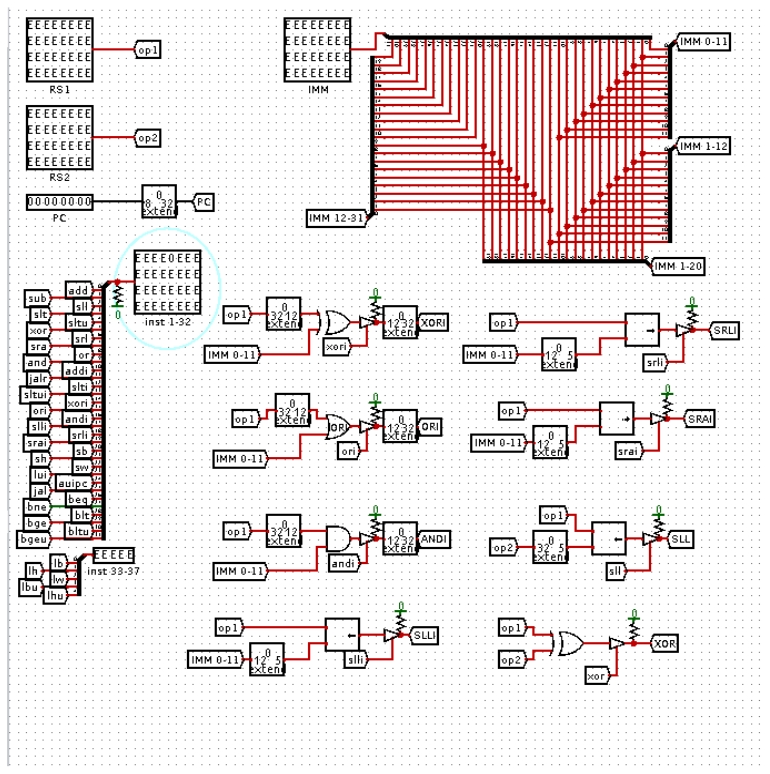}
    \caption{ALU: Part 1}
  \end{minipage}
  \hfill
  \begin{minipage}{0.6\textwidth}
    \includegraphics[width=\linewidth]{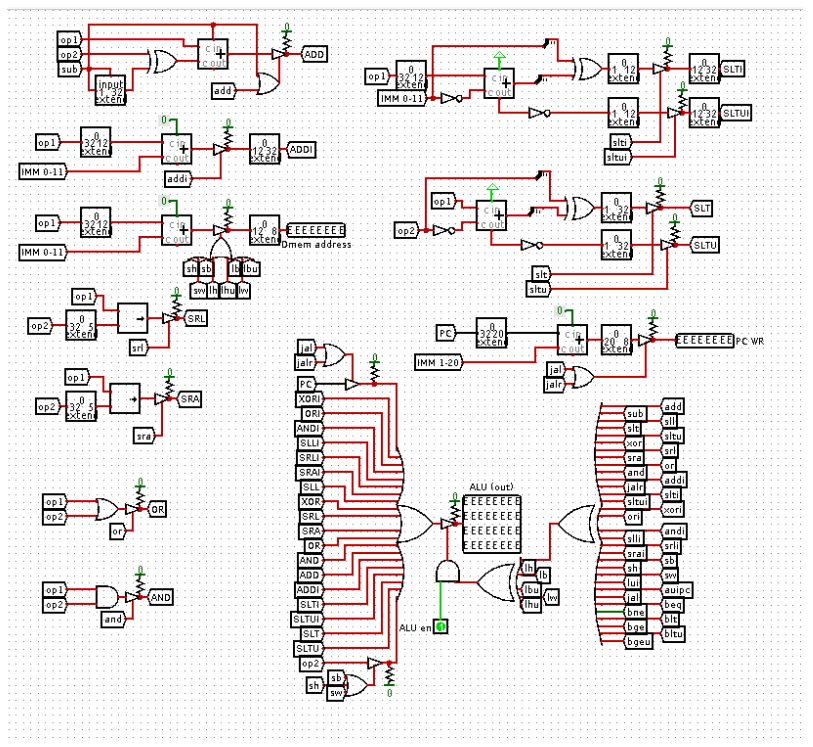}
    \caption{ALU: Part 2}
  \end{minipage}
  
  \vspace{0.5cm} 
  
  \begin{minipage}{0.25\textwidth}
    \includegraphics[width=\linewidth]{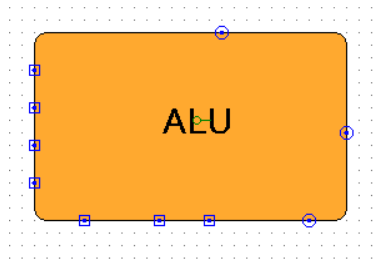}
    \caption{ALU block}
  \end{minipage}
\end{figure}

Our ALU should perform following operations:

\subsubsubsection{\textbf{Computational Instructions}}
\begin{itemize}
    \item \textbf{add, addi, sub}: Perform addition and subtraction. The immediate value in the addi
instruction is a 12-bit signed value. The sub instruction subtracts the second source
operand from the first. There is no subi instruction because addi can add a negative
immediate value.
    \item \textbf{sll, slli, srl, srli, sra, srai}: Perform logical left and right shifts (sll and srl), and arithmetic right
shifts (sra). Logical shifts insert zero bits into vacated locations. Arithmetic right shifts
replicate the sign bit into vacated locations. The number of bit positions to shift is taken
from the lowest 5 bits of the second source register or from the 5-bit immediate value.
    \item \textbf{and, andi, or, ori, xor, xori}: Perform the indicated bitwise operation on the two source
operands. Immediate operands are 12 bits.
    \item \textbf{slt, slti, sltu, sltui}: The set if less than instructions set the destination register to 1 if the first
source operand is less than the second source operand: This comparison is in terms of
two’s complement (slt) or unsigned (sltu) operands. Immediate operand values are 12 bits.
    \item \textbf{lui}: Load upper immediate. This instruction loads bits 12-31 of the destination register with
a 20-bit immediate value. Setting a register to an arbitrary 32-bit immediate value
requires two instructions: First, lui sets bits 12-31 to the upper 20 bits of the value. Then
addi adds in the lower 12 bits to form the complete 32-bit result. lui has two operands: the
destination register and the immediate value.
    \item \textbf{auipc}: Add upper immediate to PC. This instruction adds a 20-bit immediate value to the
upper 20 bits of the program counter. This instruction enables PC-relative addressing in
RISC-V. To form a complete 32-bit PC-relative address, auipc forms a partial result, then an
addi instruction adds in the lower 12 bits.
\end{itemize}

\subsubsubsection{\textbf{Control Flow Instructions}}

The conditional branching instructions perform comparisons between two registers and, based
on the result, may transfer control within the range of a signed 12-bit address offset from the
current PC. Two unconditional jump instructions are available, one of which (jalr) provides access
to the entire 32-bit address range.

\begin{itemize}

    \item  \textbf{beq, bne, blt, bltu, bge, bgeu}: Branch if equal (beq), not equal (bne), less than (blt), less
than unsigned (bltu), greater or equal (bge), or greater or equal, unsigned (bgeu). These
instructions perform the designated comparison between two registers and, if the
condition is satisfied, transfer control to the address offset provided in the 12-bit signed
immediate value.
    \item  \textbf{jal}: Jump and link. Transfer control to the PC-relative address provided in the 20-bit
signed immediate value and store the address of the next instruction (the return address)
in the destination register.
    \item  \textbf{jalr}: Jump and link, register. Compute the target address as the sum of the source
register and a signed 12- bit immediate value, then jump to that address and store the
address of the next instruction in the destination register. When preceded by the auipc
instruction, the jalr instruction can perform a PC-relative jump anywhere in the 32-bit
address space.
\end{itemize}

\subsubsubsection{\textbf{Memory Access Instructions}}

The memory access instructions transfer data between a register and a memory location. The
first operand is the register to be loaded or stored. The second is a register containing a memory
address. A signed 12-bit immediate value is added to the address in the register to produce the
final address for the load or store.
The load instructions perform sign extension for signed values or zero extension for unsigned
values. The sign or zero extension operation fills in all 32 bits in the destination register when a
smaller data value (a byte or halfword) is loaded. Unsigned loads are specified by a trailing u in
the mnemonic.

\begin{itemize}
    
    \item  \textbf{lb, lbu, lh, lhu, lw}: Load an 8-bit byte (lb), a 16-bit halfword (lh) or 32-bit word (lw) into the
destination register. For byte and halfword loads, the instruction will either sign-extend
(lb and lh) or zero-extend (lbu and lhu) to fill the 32-bit destination register. For example,
the instruction lw x1, 16(x2) loads the word at the address (x2 + 16) into register x1.
    \item  \textbf{sb, sh, sw}: Store a byte (sb), halfword (sh) or word (sw) to a memory location matching the size of the data value ALU along with ALU control unit(opcode) and control unit forms the core of processing of our
cpu. Using these instructions we can design our control unit accordingly.

\end{itemize}

\subsubsection{Decoder}
A Decoder (figure 9, 10 and 11) is a combinational logic circuit that converts binary information from the n coded
inputs to a maximum of 2n unique outputs. They are used in a wide variety of applications,
including instruction decoding, data multiplexing and data demultiplexing, seven segment
displays, and as address decoders for memory and port-mapped I/O. For our project we need
our decoder to give a specific signal.

A Decoder is made using a series of AND and NOT gates (as in figure 10) and the output can be varied using a
combination of these two gates. Our decoder should be capable to control ALU and control flow
unit. The decoder will give inputs from rs1, rs2, imm and according to the ‘high’ signal, it will
facilitate the execution of that particular operation in ALU.

\begin{figure}
    \centering
    \includegraphics[width=0.74\textwidth]{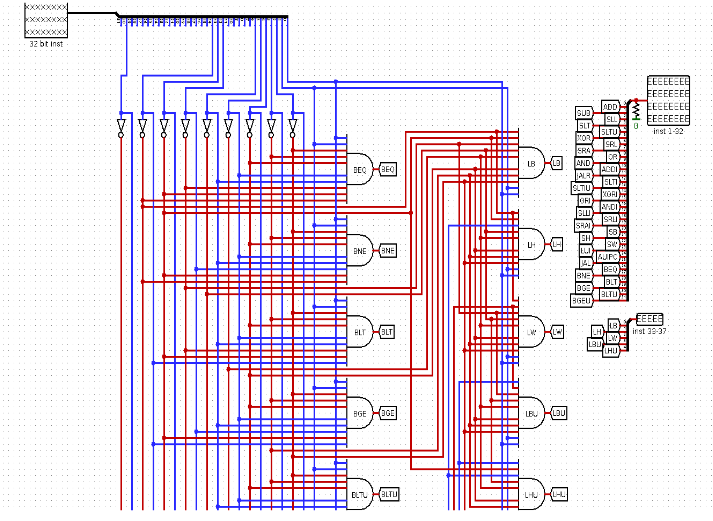}
    \caption{Basic Decoder Circuit }
    \label{fig:enter-label}
\end{figure}

\begin{figure}
    \centering
    \includegraphics[width=0.74\textwidth]{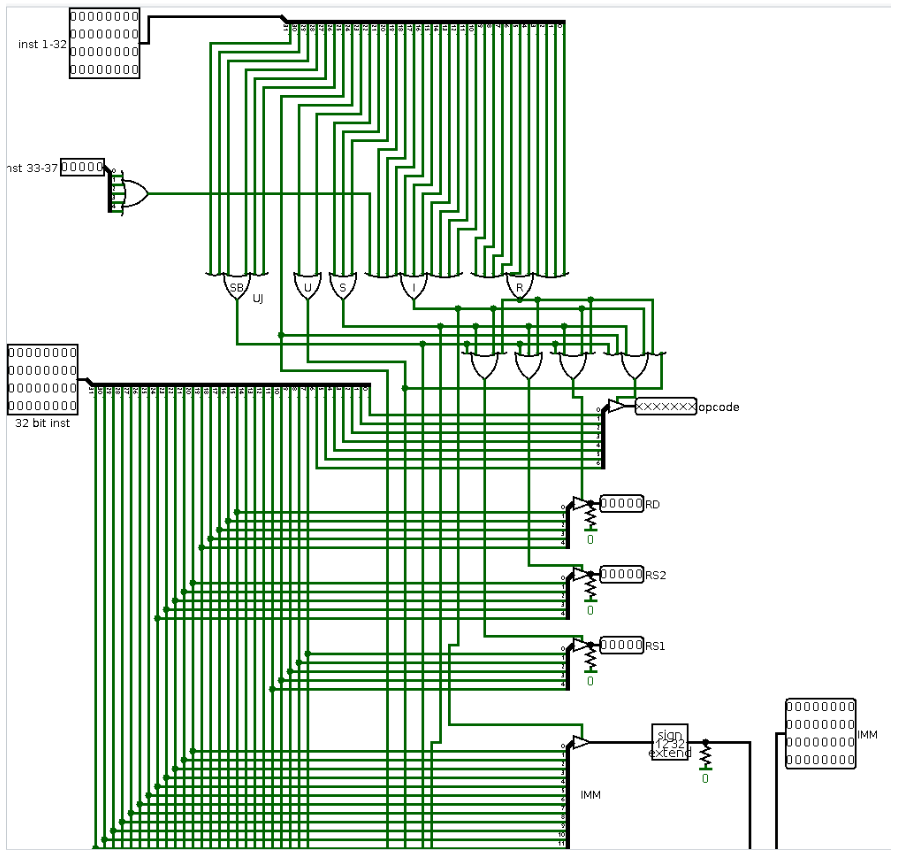}
    \caption{Decoder Circuit using a series of AND and NOT gates}
    \label{fig:enter-label}
\end{figure}

\begin{figure}
    \centering
    \includegraphics[width=0.5\textwidth]{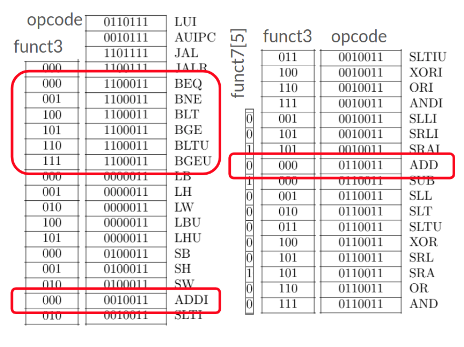}
    \caption{Instruction Decoder Table}
    \label{fig:enter-label}
\end{figure}

\subsubsection{Memory}
We use two types IMEM (fig 12) and DMEM (fig 13) for our project. IMEM stands for instruction memory where
DMEM stands for Data Memory.
IMEM can only be read by the cpu where DMEM can be read as well as written. Both these
memories are made using registers.

\begin{figure}
    \centering
    \includegraphics[width=0.88\linewidth]{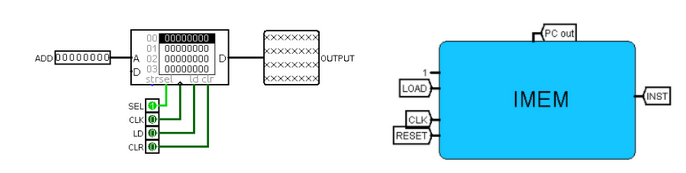}
    \caption{IMEM}
    \label{fig:enter-label}
\end{figure}

\begin{figure}
    \centering
    \includegraphics[width=0.8\linewidth]{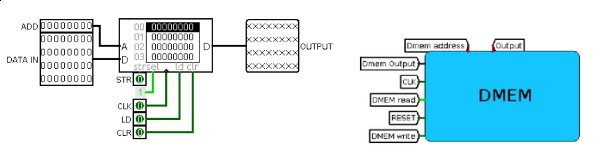}
    \caption{DMEM}
    \label{fig:enter-label}
\end{figure}

\begin{itemize}
    \item \textbf{Addressing Memory}
    
We have to provide memory with load and store instructions to read and write data. Both
load and store instructions require an address from which to read, or to which to write. As
with the IMem, this is a byte-address. Loads and stores can read/write single bytes,
half-words (2 bytes), or words (4 bytes/32 bits).The address for loads/stores is computed
based on the value from a source register and an offset value (often zero) provided as the
immediate.

\textbf{address = rs1 + imm}

\begin{figure}
    \centering
    \includegraphics[width=0.55\linewidth]{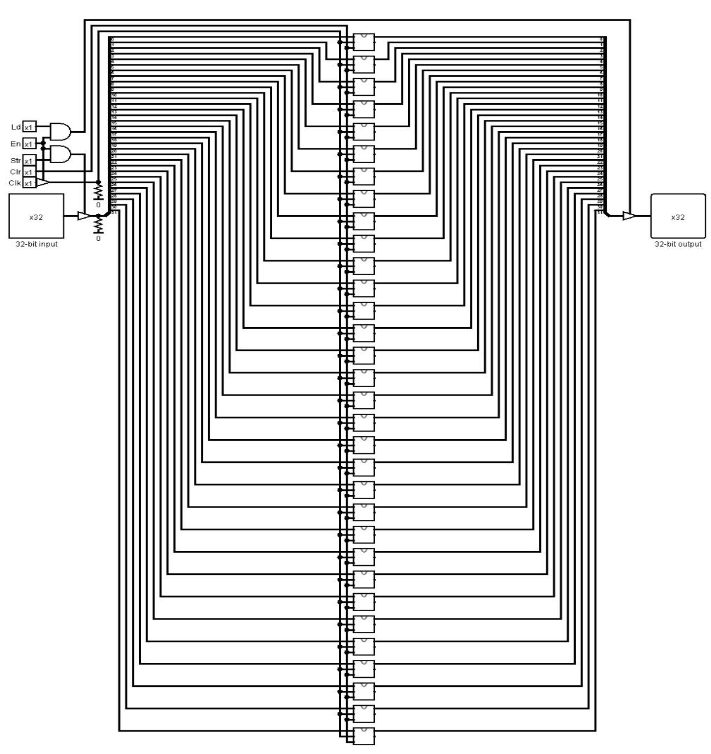}
    \caption{1 kb memory}
    \label{fig:enter-label}
\end{figure}

    \item \textbf{Load}:
    
A load instruction (LW,LH,LB,LHU,LBU) takes the form:
LOAD rd, imm(rs1). It uses the I-type instruction format: It writes its destination register
with a value read from the specified address of memory, which we can denote as: 

\textbf{rd <=DMem[addr] (where, addr = rs1 + imm)}
    \item \textbf{Stores}:
    
A store instruction (SW,SH,SB) takes the form:
STORE rs2, imm(rs1) It has its own S-type instruction format: It writes the specified
address of memory with a value from the rs2 source register: 

\textbf{DMem[addr] <= rs2 (where,addr = rs1 + imm)}
    \item \textbf{Address logic}:
    
The address computation, rs1 + imm, is the same computation performed by ADDI. Since
load/store instructions do not otherwise require the ALU, we will utilize the ALU for this
computation.

    \item \textbf{Data Memory}
    
Unlike our register file, which is capable of reading two values each cycle and, on the
same cycle, writing a value, our memory needs only to read one value or write one value
each cycle to process a load or a store instruction
\end{itemize}

For this project we will be using an inbuilt RAM module in Logisim. 

\begin{figure}
    \centering
    \includegraphics[width=0.35\linewidth]{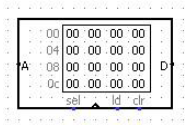}
    \caption{Inbuilt RAM module in Logisim}
    \label{fig:enter-label}
\end{figure}

The instruction memory has a single read port.1 It takes a 32-bit instruction address input, A, and
reads the 32-bit data (i.e., instruction) from that address onto the read data output, RD. The
32-element × 32-bit register file has two read ports and one write port. The read ports take 5-bit
address inputs, A1 and A2, each specifying one of 25 = 32 registers as source operands. They read the
32-bit register values onto read data outputs RD1 and RD2, respectively. The write port takes a 5-bit
address input, A3; a 32-bit write data input, WD; a write enable input, WE3; and a clock. If the write
enable is 1, the register file writes the data into the specified register on the rising edge of the
clock.The data memory has a single read/write port. If the write enable, WE, is 1, it writes data WD
into address A on the rising edge of the clock. If the write enable is 0, it reads address A onto RD

\subsubsection{Program Counter (PC)}
The PC (fig 16) is a byte address, meaning it references the first byte of an instruction in the IMem.
Instructions are 4 bytes long, so, although the PC increment is depicted as "+1" (instruction), the
actual increment must be by 4 (bytes). The lowest two PC bits must always be zero in normal
operation.  \\
The program counter (PC), commonly called the instruction pointer (IP), and sometimes
called the instruction address register (IAR), the instruction counter, or just part of the
instruction sequencer, is a processor register that indicates where a computer is in its program
sequence. \\
The PC may be a bank of binary latches, e.g. J-K latch, each one representing one bit of
the value of the PC. \\
The instruction cycle begins with a fetch, in which the CPU places the value of the PC on
the address bus to send it to the memory. The memory responds by sending the contents of
that memory location on the data bus. (This is the stored-program computer model, in which a
single memory space contains both executable instructions and ordinary data.) Following the
fetch, the CPU proceeds to execution, taking some action based on the memory contents that it
obtained. At some point in this cycle, the PC will be modified so that the next instruction
executed is a different one (typically, incremented so that the next instruction is the one starting
at the memory address immediately following the last memory location of the current
instruction). \\
Program counter is an essential part of program execution. It is also where the concept of JUMP/
GOTO gets implemented. In a raw sense a program is a set of instructions. To execute any set of
instructions we need to move from each instruction to the next. One can imagine the program
counter as an index finger as it navigates each word of a text. In its natural state it will keep
moving forward/down the instructions. But in a special case where it is told to go back or
forward to a particular word, it can directly land on the specific instruction/word specified. It can
do this in both directions: forward and backward.

\begin{itemize}
    \item Select Pin: To choose if the program counter should increment sequentially or to jump to
a given address
    \item Input Destination : The address to which the program counter should directly jump to.
    \item Clock : Increments the Pc with each cycle or jumps to the destination given as input.
    \item Clear : Resets the PC
    \item Output : The address destination where the PC points at its current state.
\end{itemize}

\begin{figure}
    \centering
    \includegraphics[width=0.85\linewidth]{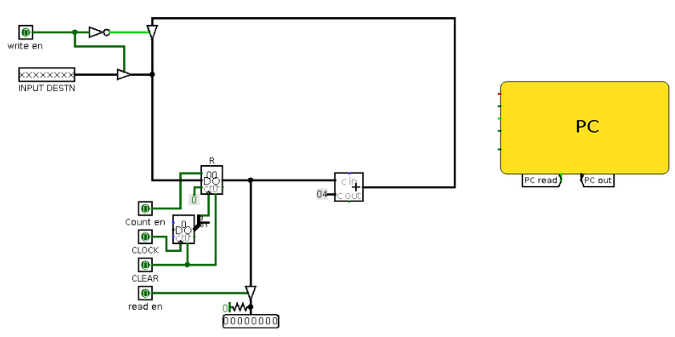}
    \caption{Program Counter}
    \label{fig:enter-label}
\end{figure}

\subsubsection{Control Logic}
Control logic (fig 17 and 18) is a separate part of ALU which handles the logical aspect of CPU and To
understand Control Logic first we need to gain a working understanding of the Control Signals.
The various components of the computer function independently.

The programs we write are implemented by enabling the control signals in an order that our
desired result is obtained. We provide ROM with 7 bits of opcode which are used to control how
we want to display and what program we want to run by controlling the pins available. (see figure 19, the B column in fig 18 is the Opcode bits.)
By virtue of Control logic we can control

\begin{figure}
    \centering
    \includegraphics[width=0.85\linewidth]{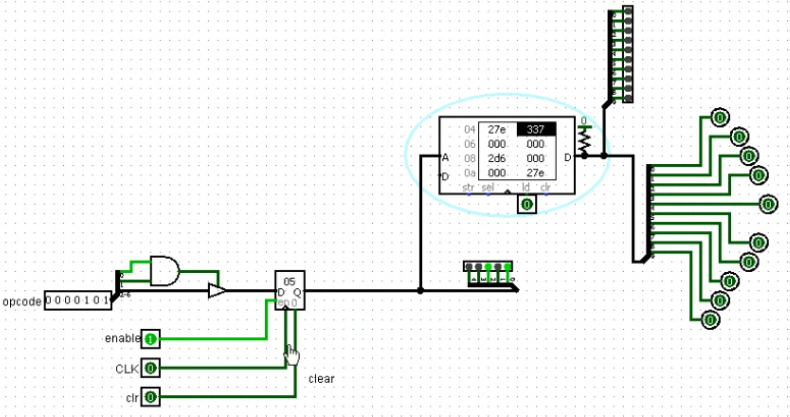}
    \caption{Control Logic (Logisim)}
    \label{fig:enter-label}

    \centering
    \includegraphics[width=0.5\linewidth]{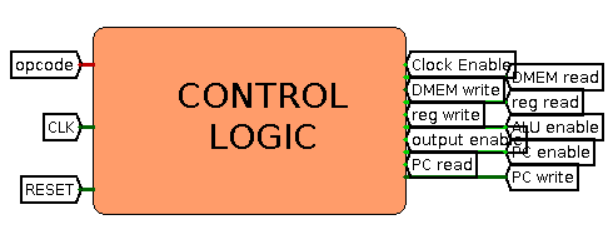}
    \caption{Control Logic Block}
    \label{fig:enter-label}

    \centering
    \includegraphics[width=0.8\linewidth]{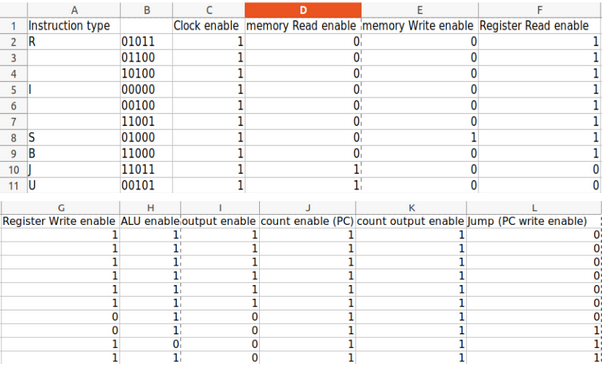}
    \caption{Control Logic}
    \label{fig:enter-label}

    \centering
    \includegraphics[width=0.8\linewidth]{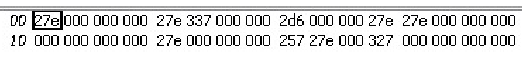}
    \caption{These hex numbers represent the current status of the 10 pins we used to control}
    \label{fig:enter-label}  
\end{figure}

\begin{itemize}
    \item Clock enable
    \item memory Read enable
    \item memory Write enable
    \item Register Read enable
    \item Register Write enable
    \item ALU enable
    \item output enable
    \item count enable (PC)
    \item count output enable
    \item Jump (PC write enable)
\end{itemize}

We are using python code to to convert binary to hex so that the program can be written
into IMEM and DMEM. (fig 20)

\subsection{Output}
Output (fig 22 and 23) will show the final result which we wish to see after running the program of our choice.
The display segment consists of binary input first going through a 8 bit decoder through
encoder in parallel connection which is then further connected to 7 segment display.
The method used to convert binary into hex numbers is “Double Dabble”. \\
\begin{figure}
    \centering
    \includegraphics[width=0.8\linewidth]{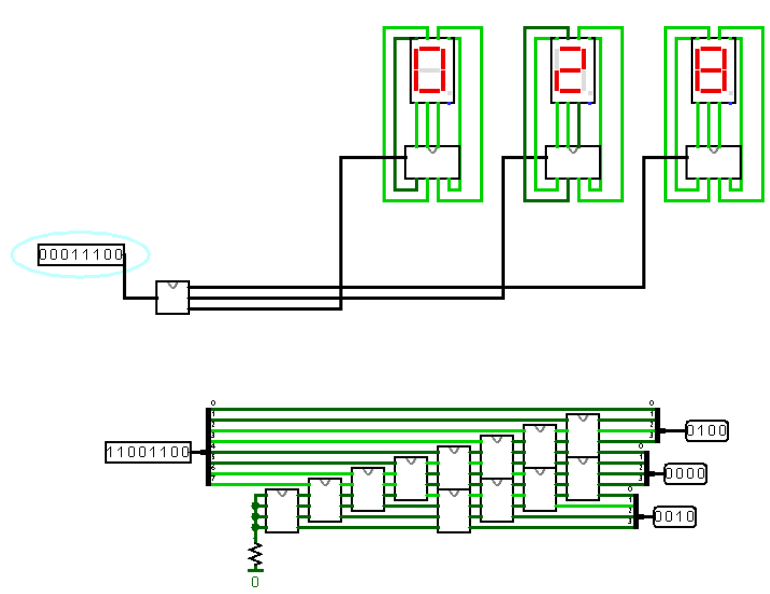}
    \caption{8 bit logic decoder}
    \label{fig:enter-label}
    
    \centering
    \includegraphics[width=0.8\linewidth]{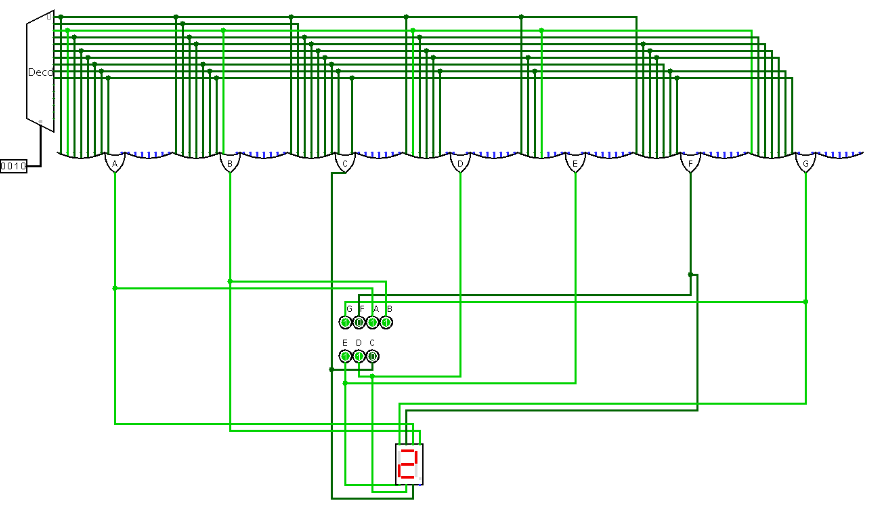}
    \caption{7 bit logic encoder}
    \label{fig:enter-label}
\end{figure}

\textbf{Double Dabble:} The double dabble algorithm is used to convert binary numbers into
binary-coded decimal (BCD) notation. It is also known as the shift-and-add-3 algorithm, and
can be implemented using a small number of gates in computer hardware, but at the expense
of high latency.

\subsection{CPU Workflow} (as per fig 23)
\begin{itemize}
    \item Program Counter (PC) gives the IMEM address where the instruction is
stored.
    \item This instruction is then identified and decoded in rs1, rs2, rd, IMM, opcode
    fields according to type of instructions.
    \item Address of rs1, rs2 and rd is fed to the Register File which gives
    corresponding data at that register.
    \item This register data is then fed to ALU which performs arithmetic operations
    and gives output.
    \item This output is either stored in DMEM or it is displayed on seven segment
    display.
    \item For load instructions, data is fetched from DMEM into registers and then it
    can be used in next instructions.
\end{itemize}

\begin{figure}
    \centering
    \includegraphics[width=0.8\linewidth]{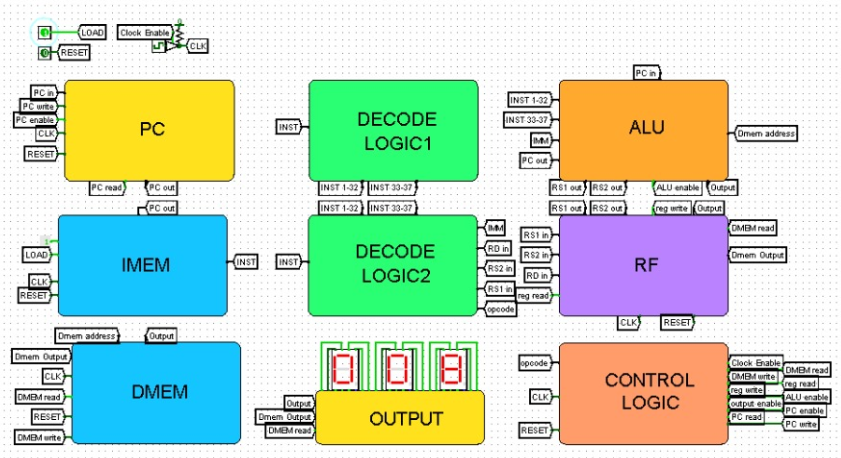}
    \caption{CPU Workflow}
    \label{fig:enter-label}
\end{figure}

\section{Conclusion}
The aim of creating a cpu core is completed. This CPU can handle basic functions like
additions, subtraction and also follows 37 RISC-V instructions. A lot was gained by
undertaking this project, majorly it deepened my understanding of Computers and
their internal workings. How memory works, how actionable currents in the circuit can
do mathematical and logical operation. We got to explore the basics of this domain and
the potential advancements that can happen in this domain. We also got to learn how
to use logisim and it function deeply in this project.
Our future goal is to add more instrcutions and functionality to this model and to
re-create this cpu using Verilog HDL.


\begin{thebibliography}{1}

  \bibitem{b1}
	Toms42. ``Logisim RISC-V CPU.'' GitHub.
	\newblock \url{https://github.com/Toms42/logisim-RISC-V-CPU}

	\bibitem{b2}
	ninja3011. ``8-bit-computer: Simulate and build Ben Eaters 8 Bit Computer.'' GitHub.
	\newblock \url{https://github.com/ninja3011/8bitcomputer}

	\bibitem{b3} 
	ninja3011. ``RISC-V pipelined core, housing a RV64I instruction set.'' GitHub.
	\newblock \url{https://github.com/ninja3011/riscv-cpu-core}

	\bibitem{b4}
	University of California, Berkeley. ``EECS Instructional and Electronics Support.''
	\newblock \url{https://inst.eecs.berkeley.edu/}

	\bibitem{b5}
	edX. ``Building a RISC-V CPU Core.''
	\newblock \url{https://learning.edx.org/course/course-v1:LinuxFoundationX+LFD111x+1T2021/home}

	\bibitem{b6}
	Eater, Ben. ``Ben Eater'' YouTube.
	\newblock \url{https://www.youtube.com/@BenEater}

 \end{thebibliography}
\end{document}